\documentclass[a4paper]{emulateapj} 

\usepackage{hyperref}
\usepackage{color}

\hypersetup{
    bookmarks=true,         
    unicode=false,          
    pdftoolbar=true,        
    pdfmenubar=true,        
    pdffitwindow=true,     
    pdfstartview={FitH},    
    pdftitle={Chronography using BHB stars},    
    pdfauthor={rsantucci},     
    pdfsubject={Astronomy},   
    pdfcreator={dvipdf},   
    pdfproducer={dvipdf}, 
    pdfkeywords={halo},
    pdfnewwindow=true,      
    colorlinks=true,       
    linkcolor=red,          
    citecolor=blue,        
    filecolor=magenta,      
    urlcolor=cyan,           
    breaklinks=true,
    linktocpage
}

\newcommand{\metal}{[Fe/{}H]}

\newcommand{\logg}{log\,$g$\,}

\shorttitle{Chronography of the Galactic Halo}
\shortauthors{Santucci et al.}

\begin{document}

\title{Chronography of the Milky Way's Halo System with Field \\ Blue Horizontal-Branch Stars}

\author{Rafael M. Santucci}
\affil{Departamento de Astronomia - Instituto de Astronomia,
Geof\'isica e Ci\^encias Atmosf\'ericas, \\ Universidade de S\~ao Paulo,
S\~ao Paulo, SP 05508-900, Brazil}
\email{rafaelsantucci@usp.br}

\author{Timothy C. Beers, Vinicius M. Placco, Daniela Carollo}
\affil{Department of Physics and JINA Center for the Evolution of the Elements, \\
University of Notre Dame, 225 Nieuwland Science Hall, Notre Dame, IN 46556, USA}
\email{tbeers@nd.edu}
\email{vplacco@nd.edu}
\email{dcaroll1@nd.edu}

\author{Silvia Rossi}
\affil{Departamento de Astronomia - Instituto de Astronomia,
Geof\'isica e Ci\^encias Atmosf\'ericas, \\ Universidade de S\~ao Paulo,
S\~ao Paulo, SP 05508-900, Brazil}
\email{rossi@astro.iag.usp.br}

\author{Young Sun Lee}
\affil{Department of Astronomy and Space Science, Chungnam National 
University, Daejeon 34134, Republic of Korea}
\email{youngsun@cnu.ac.kr}

\author{Pavel Denissenkov}
\affil{Department of Physics \& Astronomy, University of Victoria,
Victoria, BC, V8W3P6, Canada}
\email{pavelden@uvic.ca}

\author{Jason Tumlinson}
\affil{Space Telescope Science Institute, Baltimore, MD 21218, USA}
\email{tumlinson@stsci.edu}

\author{Patricia B. Tissera}
\affil{Departamento de Ciencias Fisicas and Millennium Institute of
Astrophysics, \\ Universidad Andres Bello, Av. Republica 220, Santiago,
Chile} 
\email{patricia.tissera@unab.cl}

\begin{abstract}

In a pioneering effort, Preston et al. reported that the colors of blue
horizontal-branch (BHB) stars in the halo of the Galaxy shift with
distance, from regions near the Galactic center to about 12 kpc away, and
interpreted this as a correlated variation in the ages of halo stars,
from older to younger, spanning a range of a few Gyrs. We have applied
this approach to a sample of some 4700 spectroscopically confirmed BHB
stars selected from the Sloan Digital Sky Survey to produce the first
``chronographic map'' of the halo of the Galaxy. We demonstrate that the
mean de-reddened $g-r$ color, $\langle$($g-r$)$_0\rangle$, increases
outward in the Galaxy from $-0.22$ to $-0.08$ (over a color window
spanning [$-0.3:0.0$]) from regions close to the Galactic center to
$\sim$40 kpc, independent of the metallicity of the stars. Models of the
expected shift in the color of the field BHB stars based on modern
stellar evolutionary codes confirm that this color gradient can be
associated with an age difference of roughly 2-2.5 Gyrs, with the oldest
stars concentrated in the central $\sim 15$ kpc of the Galaxy. Within
this central region, the age difference spans a mean color range of
about 0.05 mag ($\sim$ 0.8 Gyrs). Furthermore, we show that
chronographic maps can be used to identify individual substructures,
such as the Sagittarius Stream, and overdensities in the direction of
Virgo and Monoceros, based on the observed contrast in their mean BHB
colors with respect to the foreground/background field population.

\end{abstract}


\section{Introduction} 
\label{c1}

Study of the stellar populations in the Galaxy has long been a pillar of
the effort to understand its formation and evolution, ever since the
introduction of the concept in the middle of the last century by
\citet{baade54}. Indeed, the existence (or not) of gradients in the age
and/or metallicity of the globular cluster system was a central piece of
evidence used by \citet{Searle78} to support their suggestion for a
hierarchical assembly of the Galaxy, contrasting with the rapid,
monolithic collapse model put forward by \citet{Eggen62}. More recently, 
\citet{Lee1994} argued in support of age as the primary ``second parameter''
in the morphology of cluster horizontal-branches, concluding that
globular clusters located within 15-20 kpc of the Galactic center are on
the order of 2 Gyrs older than clusters located outside this region.
A modern discussion of the globular-cluster age-metallicity relationship (based
on refined estimates of metallicity, and in particular, age), and its
variation with distance in the Galaxy, can be found in \citet{Leaman13}.

While the globular clusters provide clear insight, it would be of even
greater importance if a similar exercise could be carried out making use
of the far more numerous field stars in the halo system. Even though
astero-seismology is making great strides toward this goal 
\citep[e.g.,][]{Soderblom13,SA15}, age determinations for
individual field stars remain a challenge, and for the present are still
quite limited in the numbers of objects (and distance ranges) that have
been studied.

The seminal work of \citet{Preston91} explored using the colors of field
blue horizontal-branch (BHB) stars as tracers of the age of the
underlying stellar population of the halo. These authors employed a
sample of some 500 BHB candidates selected from the HK objective-prism
survey \citep[][]{Beers88,Preston91b}, with available broadband $U-B$
and $B-V$ colors that distinguished them from potential contamination by
high-gravity blue stragglers or A-type main-sequence stars. Their
analysis indicated that there exists a shift in the mean BHB colors,
$\langle (B-V)_0 \rangle$, on the order of 0.025 ($\pm$0.004) mag with
Galactocentric distance over the region 2 kpc $< R <$ 12 kpc, which they
interpreted as an age gradient in the inner portion of the Galactic halo
of $\sim$ 2 Gyr, similar to that suggested by \citet{Zinn80} to exist for
inner-halo globular clusters.

In this letter we revisit the issue of the mean colors of halo BHB
stars, and their possible variation with Galactocentric distance, making
use of a much larger sample of field stars than were available to
Preston et al. Our sample is drawn from photometry obtained by the Sloan
Digital Sky Survey \citep[SDSS;][]{York00}, and with available SDSS
spectroscopy that allows us to confidently differentiate BHB stars from
possible contaminants, such as blue stragglers. This sample is used to
produce, for the first time, a chronographic ``age map'' of the
underlying stellar population of the halo system of the Galaxy. This map
not only confirms the previous result from \citet{Preston91}, albeit
with a somewhat smaller age range found for the central region of the
Galaxy, but permits the identification of stellar debris from the
Sagittarius Stream, as well as the Virgo and Monoceros
overdensities.

\begin{figure}[!ht]
\epsscale{1.0} 
\plotone{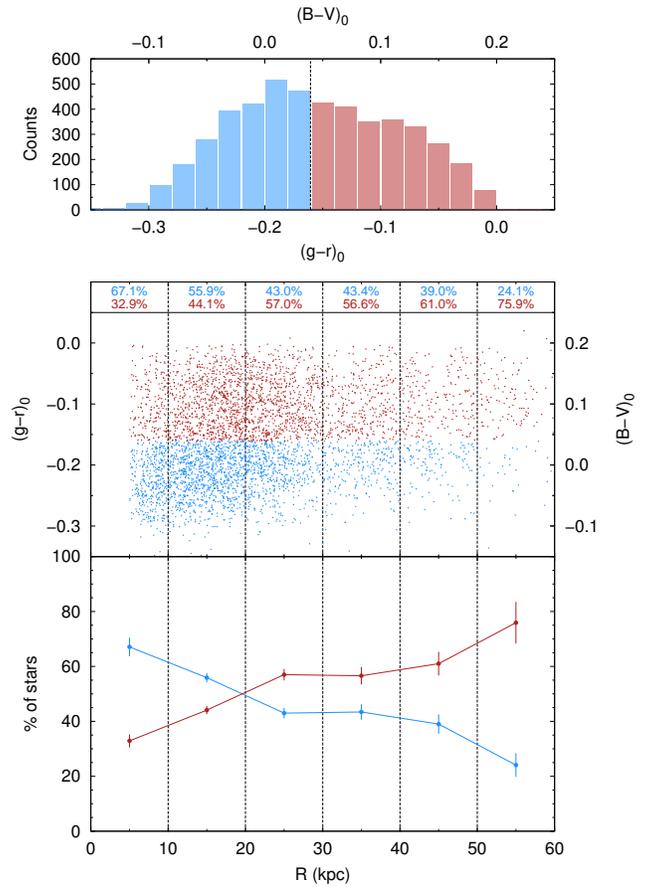}
\caption{{\it Top panel}: Distributions of $(g-r)_0$ and $(B-V)_0$
colors for the BHB stars in our sample. The vertical black dashed line marks the
median $(g-r)_0$ color ($-0.161$). {\it Middle panel}: $(g-r)_0$ as a
function of Galactocentric distance, $R$, for our sample, color-coded
according to whether the stars lie to the blue side (blue dots) or red
side (red dots) of the median color. {\it Bottom panel}: Fractions of stars in the blue and red
color windows, as a function of $R$, in bins of width 10 kpc, indicated
by the blue and red lines. Poisson error bars are shown for each point.}
\label{Fig1}
\end{figure}

\section{The BHB Sample}
\label{c2}

Our BHB sample is obtained from stars with photometry and spectroscopy
released as part of SDSS DR8 \citep{Aihara11}. The adopted restrictions 
for the selection of field BHB stars are fully described in
\citet{Santucci15}. Briefly, stars were required to satisfy: (1) A color
cut: $0.60 < (u-g)_0 < 1.60$ and $-0.50 < (g-r)_0 < 0.05$\footnote{All
magnitudes and colors presented in this work have been corrected for 
interstellar absorption and reddening according to \citet{Schlegel98}.} and (2) Surface
gravities, \logg{}, as derived from application of a recent version of
the SEGUE Stellar Parameter Pipeline (SSPP; see \citealt{Lee08a,Lee08b,
Allende08,Smolinski11,Lee11} for details): 3.0 $<$ \logg{} $<$ 3.8. For 
a discussion of the \logg\ uncertainties and the application of surface-gravity estimates 
for selection of BHB stars, see \citet{Lee08b} and \citet{Santucci15}, respectively.

\begin{figure}[!ht]
\epsscale{1.22} 
\plotone{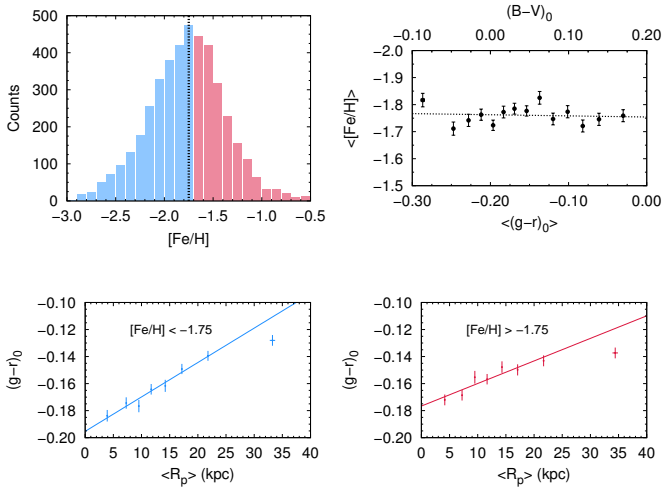}
\caption{{\it Top-left panel}: Distribution of metallicity, \metal, for
the BHB sample. The distribution is divided into two equal parts 
(at \metal$=-1.75$), color-coded according to \metal{}. {\it Top-right panel}: 
The mean \metal as a function of mean $(g-r)_0$ color over the full extent 
of our selected color window, $-0.30 < (g-r)_0 < 0.00$. There is no 
variation of $\langle$\metal$\rangle$ with $\langle$ $(g-r)_0$ $\rangle$. 
{\it Lower panels}: Robust fits to $(g-r)_0$ vs. $R_p$ for both metallicity intervals, 
color-coded according to the distribution shown in the top-left panel.
Note that the color shift with distance is similar for both metallicity
intervals, but the higher interval exhibits a slightly shallower slope.
The points in these panels represent an average of 300 BHBs in each bin;
error bars are the standard errors in the mean. The outer-most point in both panels
is located at distances where the contribution from resolved substructures becomes
substantial, giving rise to the apparent deviation.}
\label{Fig2}
\end{figure}

These restrictions resulted in an initial sample of $\sim$4800 BHB stars.
For the color window we employ for this study, $-0.3 < (g-r)_0 < 0.0$,
there are some 4700 BHB stars. In order to carry out an approximate
comparison with the \citet{Preston91} results, we also obtained a
transformation to the Johnson system, based on \citet{Zhao06}. See
\citet{Beers12} for comments on the range of validity of these
transformations.

\begin{equation} 
{\rm V_0} = g - 0.561 \cdot (g-r)_0 - 0.004, 
\label{eqV}
\end{equation} 
\noindent and 
\begin{equation} (B-V)_0 = 0.916 \cdot (g-r)_0 + 0.187.
\label{eqBV} 
\end{equation}

Distances were estimated adopting the absolute magnitude calibration described 
by \citet{Deason11b}: 

\begin{eqnarray} 
{\rm {M_{g}}}= 0.434 - 0.169 \cdot (g - r)_0 + 2.319 \cdot {(g- r)_0}^2 \nonumber \\ 
+ 20.449 \cdot {(g - r)_0}^3 + 94.517 \cdot {(g - r)_0}^4, 
\label{eqMg} 
\end{eqnarray}

\noindent Note that \citet{Santucci15} found little difference between
the above relationship and those that included a metallicity-dependent
term, at least for stars in our metallicity range. 

Figure~\ref{Fig1} (top panel) shows the distributions of $(g-r)_0$
colors for our sample stars (an approximate $(B-V)_0$ scale is shown
at the top of the panel). The vertical black dashed line in this panel
marks the median $(g-r)_0$ value of the distribution 
$(g-r)_0 = -0.161$), which was used to split the BHB colors into two pieces,
which we refer to as the red-window (RW) and blue-window (BW) stars. The
lower panels show the distribution of the BW and RW stars as a function
of Galactocentric distance, $R$. It is immediately clear that the BW BHB
stars are preferably found closer to the Galactic center, while the RW
BHB stars are found over a wider range in distance.\footnote{This result
is reminiscent of Figure 1 of \citet{Preston91}, which showed the
distribution of apparent $V_0$ magnitude and ($B-V$)$_0$ colors for
their sample stars in three angular ranges toward the Galactic center.}
The crossover point occurs at around $R =$ 20 kpc, beyond which there
are more RW BHB stars then BW BHB stars. The percentages of BW and RW
BHB stars in each $R-$step are shown at the top of the lower panel, and
are indicated by blue and red lines, respectively.

The \citet{Preston91} color window, $-0.02 < (B-V)_0 < 0.18$ 
($-0.23 < (g-r)_0 <-0.01$) is narrower than our present window, 
which covers $-0.10 < (B-V)_0 < 0.20$ ($-0.30 <  (g-r)_0 < 0.00$). We chose this
``extended color window'' to include bluer stars in order to increase
the dynamical range of the color shift, since we can confidently
classify our targets as BHB stars on the basis of spectroscopic
measurements. We have verified that we do not undersample these bluer
stars (due to their intrinsically lower luminosities compared with the
redder stars) at the apparent magnitude corresponding to the maximum
distances explored by our sample (for $g_0~<$ 20, the blue stars
do not begin to fall out of the sample, relative to the redder BHB
stars, until distances beyond 50 kpc are reached).

\section{The Variation of BHB Colors with Metallicity}
 
\citet{Preston91} demonstrated to their satisfaction that metallicity
was not responsible for the observed color shift of their BHB
candidates, but this was in the absence of spectroscopic determinations
for their entire sample. Since we have that information for all of the
stars in our sample, we can examine this question more quantitatively.
The top-left panel of Figure \ref{Fig2} shows the distribution of
spectroscopically determined metallicity, \metal{}, for our sample. The
top-right panel shows the mean metallicity $\langle$\metal$\rangle$, as
a function of the mean color, $\langle$($g-r$)$_0\rangle$. As is clear
from inspection, there is no variation.    

As a further check, we have divided this distribution into two equal
parts at the median value of \metal $=-$1.75. The lower panels of this
figure show the distribution of mean $(g-r)_0$ colors for each of these
metallicity regions, as a function of projected Galactocentric distance,
$R_p$. The color shift is present for both intervals, but it is slightly
shallower for the relatively more metal-rich subset of the data.

\section{The Observed Color Shift for Halo BHB Stars and the Derived Chronographic Map}
\label{c3}

\begin{figure*}[!ht]
\epsscale{1.20} 
\plotone{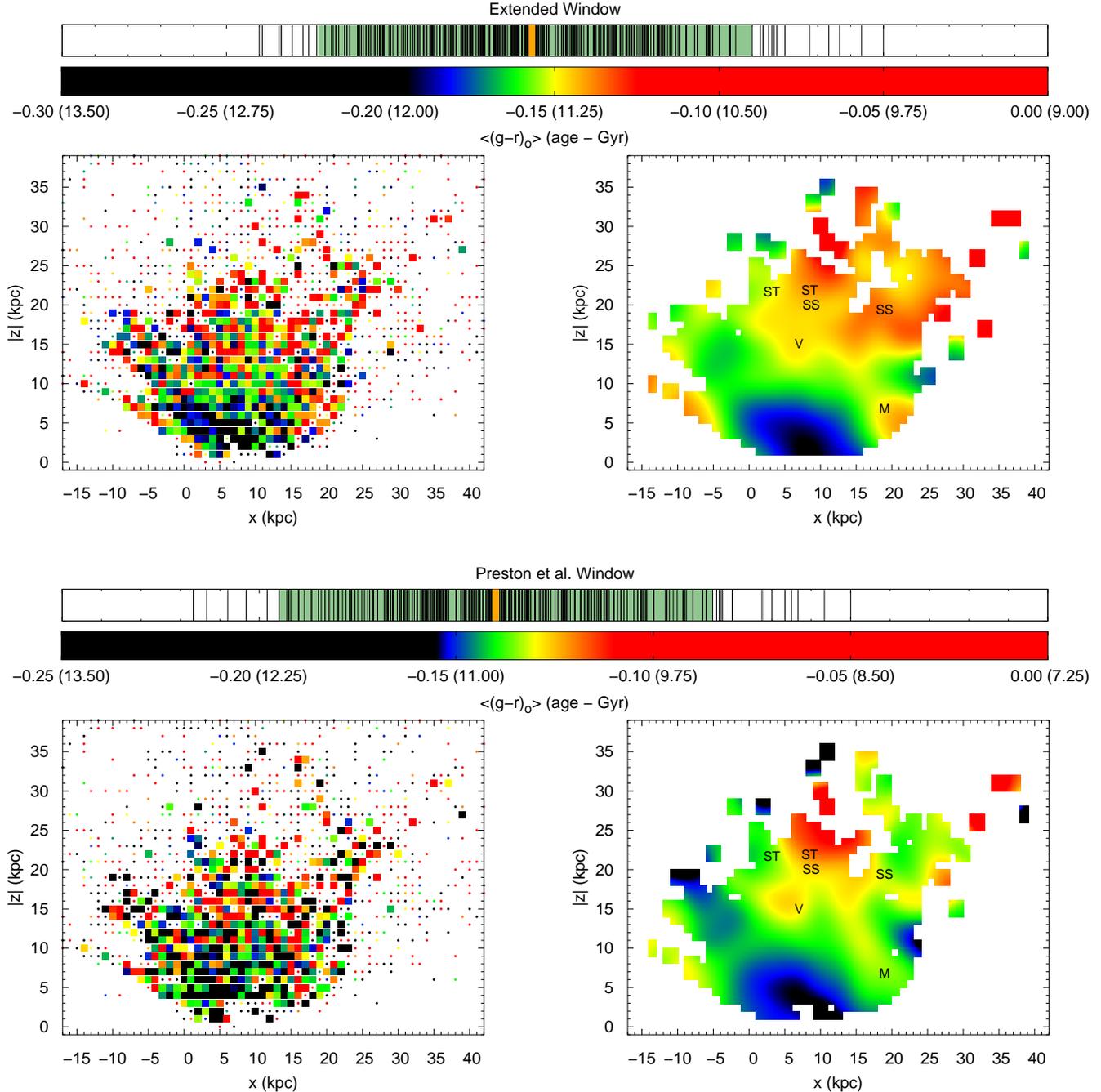}
\caption{Color (age) maps for the Extended Window (top row) and
\citet{Preston91} Window (bottom row). {\it Left panels}: Distribution of data 
in the X vs. $|{\rm {Z}}|$ plane, in a square grid with 1 kpc pixels. The filled
squares represent pixels with at least 3 BHB stars present (accepted
pixels); the color represents the mean $(g - r)_0$ for each
pixel. The filled dots are pixels with only one star. {\it Right
panels}: Smoothed version of the left panels, taking into account only
the accepted pixels. The numbers in parentheses in the color scale are
the corresponding ages in Gyrs. The distributions above the color scale are
stripe plots of the mean $(g - r)_0$ colors for the accepted pixels. The
orange bars represents the medians, and the green shaded areas indicate
the $\pm$ 2-sigma regions. The labels ST, SS, V and M denote the location
of known major streams and overdensities, such us the Sagittarius (Sgr)
Stream, the Virgo Over Density (VOD; V), and Monoceros (M). The ST
labels indicate portions of the trailing arm of Sgr in the North Galactic
Hemisphere, while the SS labels identify portions of the Sgr Stream
located in the South Galactic Hemisphere, and includes contributions
from both the leading and trailing arms. The VOD extends farther north
of the labelled location in the map. A number of low-contrast features
are present in these maps without clear associations with known
structure, although some could be numerical artifacts due to the limited
number of stars in the sample. The distance of the Sun from the Galactic
center for these maps is assumed to be 8.5 kpc.
\\
\\ }
\label{Fig3}
\end{figure*}

Figure~\ref{Fig3} is a map of the distribution of the $(g-r)_0$ colors
of halo-system BHB stars in the plane X vs. $|$Z$|$. For comparison,
this figure shows maps based on both the extended color window we employ
(upper panel) and the original Preston et al. window (lower panel). As
can be appreciated from inspection of these maps, the results are quite
similar.

\citet{Preston91} proposed that the observed shift in the mean colors of
BHB stars in their sample arose due to changes in the age of the
underlying stellar population, and obtained an approximate relationship
between the observed color shift and the proposed age shift\footnote{$\Delta
t_{9}~=~-80(\pm 40)\cdot\Delta\langle{}B-V{}\rangle_{w}$, where
$\Delta$t$_{9}$ is the age variation in Gyr and 
$\Delta\langle{}B-V\rangle_{w}$ is the window of mean color over which
the variation occurs.}, using stellar-evolution models that were
then available, and an empirical calibration based on globular
cluster horizontal branches. 

\begin{figure}[!ht]
\epsscale{1.22} 
\plotone{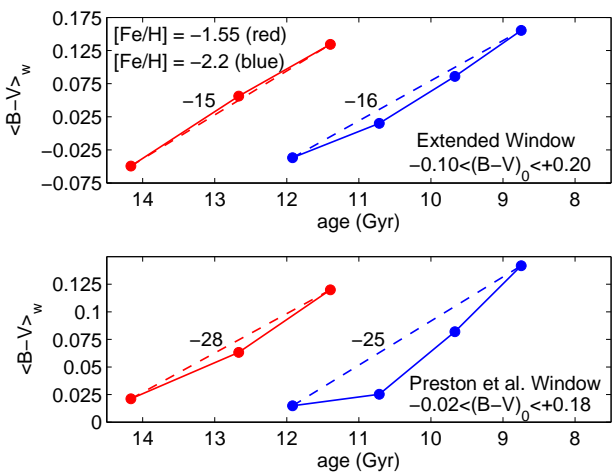}
\caption{The dependencies of $\langle B-V\rangle_w$ on the HB age for 
two metallicities (indicated in the upper panel) estimated from our HB 
population-synthesis simulations (filled circles connected by solid lines).
Dashed lines and the numbers next to them give the mean slopes 
($\Delta t_9/\Delta\langle B-V\rangle_w$) of the dependencies. The results 
are shown for the extended window (upper panel) and the original Preston
et al. window (lower panel).}
\label{fig:PavelFig}
\end{figure}

We have reconsidered the approximation derived by Preston et al., using
revision 5329 of the MESA stellar-evolution code \citep{paxton:11},
tuned to produce evolutionary tracks and zero-age horizontal-branch
(ZAHB) models close to those calculated with the Victoria code 
\citep{vandenberg:12}. The position of a ZAHB star on the color-magnitude
diagram, e.g., relative to the selected color window, is determined by
its initial mass, metallicity, \metal{}, initial helium ($Y$) and CNO
abundances, as well as by the amount of mass lost on the red giant
branch. For simplicity, we have assumed that the stars have the same
$Y=0.25$ and [$\alpha$/Fe] = +0.4. To further reduce the number of free
parameters, we have used the Reimers prescription \citep{reimers:77} for
RGB mass loss with $\eta = 0.4$. This yields mean ZAHB masses ($\langle
M_\mathrm{ZAHB}\rangle$) of stars with different initial masses and
metallicities for a range of horizontal-branch (HB) ages. To populate a
HB by stars with ZAHB masses close to the mean one, we have used our new
HB population-synthesis tool. It assumes that the ZAHB masses are
normally distributed around $\langle M_\mathrm{ZAHB}\rangle$ with a
dispersion $\sigma_M = 0.015\,M_\odot$, the value of which does not
affect the final results of this study. Our HB population-synthesis
simulations enable the estimation of $\langle B-V\rangle_w$ as a
function of the HB age and metallicity and, as a result, to find the
slopes $\Delta t_9/\Delta\langle B-V\rangle_w$ (Figure
\ref{fig:PavelFig}).

For our current purpose, the zero point of the mapping from color to age
was set at 13.5 Gyr, which assumes that star formation began shortly
after the Big Bang, consistent with the recent age determination of the
very metal-poor subgiant HD~140283 \citep{vandenberg2014}. Note that
$\Delta\langle{}g-r\rangle_{w}$ is similar, though not identical, to
$\Delta\langle{}B-V\rangle_{w}$ ($\Delta\langle{}g-r\rangle_{w}$ =
$1.09 \cdot \Delta\langle{}B-V\rangle_{w}$). In order to provide an
appropriate color to age conversion, we use an average of the slopes
shown in Figure~\ref{fig:PavelFig} (Extended Window: $\Delta
t_9/\Delta\langle B-V\rangle_w = -15.5$; Preston et al. Window: $\Delta
t_9/\Delta\langle B-V\rangle_w = -26.5$). 

There are a number of distinctive features seen in the chronographic 
maps of Figure~\ref{Fig3}.  The most obvious is the concentration
of the oldest stars ($\sim 11.5$-12.5 Gyrs) within 15 kpc from the
Galactic center.  This ``chronographic sphere'' occupies the 
predicted locations for the most ancient stars in the Milky Way, as
shown by a number of modern numerical simulations of galaxy assembly
(see \S~\ref{c5}). The color difference across this
feature, $\Delta\langle{}g-r\rangle_{w}\sim 0.05$ mag, corresponds to
an age difference of about 0.8 Gyrs, a factor of $\sim$ 2.5 smaller than
that inferred previously by \citet{Preston91}, primarily due to the
revised slopes of the color-age relationship. It is worth noting that
the ancient chronographic sphere reaches well into the Solar
Neighborhood, suggesting that there should be numerous examples of
relatively nearby stars whose origin is associated with the outer-halo
population argued to be present by \citet{carollo07,carollo10},
\citet{Beers12}, and \citet{An13,An15}. Other younger features
(those with yellow and red colors in the maps) can be associated with
the Sgr Stream, and overdensities in the directions of Virgo and
Monoceros, although further interpretation is limited by the small sample
size.  

\section{Interpretation}
\label{c5}

An age gradient within the Galactic halo that places older stars nearer
the center of the Galaxy is consistent with the generic finding from
structure formation theory in a $\Lambda$CDM cosmology that the oldest
progenitors of a halo are centrally concentrated. This concentration
results from the ``inside out'' formation of $\Lambda$CDM halos, in
which any halo's central regions are built from the first progenitor
halos to collapse, while its outskirts are built preferentially from
subhalos that themselves virialized and merged at later times
\citep{White00}. This robust behavior, seen in DM-only simulations of
$\Lambda$CDM halos, need not translate directly into gradients in the
properties of stellar populations formed in these progenitor halos, as
variations in star-formation histories, chemical abundances, and stellar
physics could in principle erase any ``age'' gradient in the underlying DM
subhalos. 

However, detailed simulations of subhalos accreting over time to form a
Milky Way (MW)-sized parent halo at $z = 0$ \citep{Bullock05,Cooper10,
Tumlinson10} recover gradients in age from place to place in the
Galactic halo, with the oldest stars at a given metallicity being
preferentially concentrated at smaller radii and on more tightly bound
orbits. These simulations typically find ages of $\sim $8-12 Gyr for
stellar populations stripped from accreted halos to build the metal-poor
MW halo near the Sun \citep{Font06,Cooper10}. These ages and the
metallicities of the model populations match well with the range of ages
and metallicities in the BHB sample. While these models qualitatively
agree with our chronographic map, none of them analyzed gradients over
the same range of metallicity and age as the BHB sample, so a direct
comparison between model and data is precluded. 

The models further show that the gradients occur in some stochastic
halo-building histories, but are less strong or absent in others. This
finding hints that clear age gradients may not be generic to all
possible ways of building MW-like halos, but may arise from
particular star-formation and/or chemical enrichment conditions that
were met in our own Galaxy. Hydrodynamical cosmological simulations that
include chemical evolution show the inner regions of MW-sized
haloes to be populated by a combination of disrupted stars from accreted
satellites, as well as stars formed in situ \citep{mcCarthy2012,
tissera2012}. More detailed studies by \citet{tissera2013} found that,
while disrupted stars covered a larger range in binding energy than the
in situ stars, both stellar populations are found to be low-metallicity,
$\alpha$-element enhanced, and have ages older than $\sim 8$ Gyr, with a
mean age of $\sim11$-12 Gyr. Their findings suggest that the existence
of an age gradient could be linked to properties of the accreted
satellites in early stages of the halo assembly, and modulated by the
contribution of the in situ stars. In either case, our chronographic map
enables important tests of hierarchical models and the physical
ingredients that go into them.

\section{Conclusions}
\label{c6}

We have presented the first chronographic map of the age distribution of
stars in the halo system of the Galaxy, based on the mean colors of a
spectroscopically confirmed sample of $\sim$4700 stars from SDSS. These
results confirm and extend the previous work of \cite{Preston91}, and
are consistent with the reported age differences of globular clusters in
the inner- vs. outer-halo regions, as well as with the predictions of
current models of hierarchical galaxy assembly. 

Although it should be cautioned that variations in the colors of BHB
stars could also arise from differences in their helium and/or CNO
abundances, it is difficult to imagine why these should correlate with
distance from the Galactic center. Assuming age is indeed the primary
factor affecting the mean colors of BHB stars, this provides a powerful
new tool for the exploration of the star-formation history of halo stars
and within individual accreted structures. 

We look forward to application of the chronographic method to the full
photometric sample of BHB stars from SDSS (Carollo et al., in prep.), as
well as to other samples of resolved stellar populations in the Local
Group with suitable available data. Comparison with insights from
the Gaia mission \citep{perryman01,jordi15}, as well as application to the photometry
gathered by LSST in more distant regions of the Galaxy, and the
opportunity to identify overdensities and streams based on age contrast
rather than exclusively number density contrast, are all exciting
prospects.

\acknowledgements 

The authors thank Don Vandenberg for his thoughts and suggestions 
during the course of this work. R.M.S. and S.R. acknowledge CAPES (PROEX), CNPq, 
PRPG/USP, FAPESP and INCT-A funding. T.C.B., V.M.P., D.C. and P.D. acknowledge 
partial support for this work from grant PHY 14-30152; Physics Frontier 
Center/JINA Center for the Evolution of the Elements (JINA-CEE), awarded 
by the US National Science Foundation. Y.S.L. acknowledges support provided 
by the National Research Foundation of Korea to the Center for Galaxy Evolution 
Research (No. 2010-0027910) and partial support from the Basic Science 
Research Program through the National Research Foundation of Korea (NRF) 
funded by the Ministry of Science, ICT \& Future Planning (NRF-2015R1C1A1A02036658).
P.D. also acknowledges partial funding from a Natural Sciences and Engineering
Research Council of Canada grant to Don VandenBerg. PBT acknowledges
partial support from PICT-959-2011, Fondecyt-113350 and MUN-UNAB projects.


\begin{thebibliography}{39}
\expandafter\ifx\csname natexlab\endcsname\relax\def\natexlab#1{#1}\fi

\bibitem[{{Aihara} {et~al.}(2011){Aihara}, {Allende Prieto}, {An}}]{Aihara11}
{Aihara}, H., {Allende Prieto}, C., {An}, D. {\it{et}}~{\it{al}}. 2011, \apjs, 193, 29

\bibitem[{{Allende Prieto} {et~al.}(2008){Allende Prieto}, {Sivarani}, {Beers}}]{Allende08}
{Allende Prieto}, C., {Sivarani}, T., {Beers}, T.~C. {\it{et}}~{\it{al}}. 2008, \aj,
  136, 2070

\bibitem[An et al.(2013)]{An13} An, D., Beers, T. C., Johnson, J. A., et al. 2013, \apj, 763, 65 

\bibitem[An et al.(2015)]{An15} An, D., Beers, T. C., Santucci, R. M., et al. 2015, \apjl, submitted

\bibitem[Baade(1954)]{baade54} Baade, W. 1954, Transactions of the International
Astronomical Union VIII (Cambridge, 1954), 8, 682

\bibitem[{{Beers} {et~al.}(2012){Beers}, {Carollo}, {Ivezi{\'c}}}]{Beers12}
{Beers}, T.~C., {Carollo}, D., {Ivezi{\'c}}, {\v Z}. {\it{et}}~{\it{al}}.
  2012, \apj, 746, 34

\bibitem[Beers et al.(1988)]{Beers88} Beers, T.C., Preston, G.W., \& Shectman, S.A. 1988, \apjs, 67, 461 

\bibitem[Bullock \& Johnston(2005)]{Bullock05} Bullock, J.~S., \& Johnston, K.~V.\ 2005, \apj, 635, 931 

\bibitem[Carollo et al.(2007)]{carollo07} Carollo, D., Beers, T. C., Lee, Y. S., et al. 2007, Nature, 450, 1020

\bibitem[Carollo et al.(2010)]{carollo10} Carollo, D., Beers, T. C., Chiba, M., et al. 2010, ApJ, 712, 692

\bibitem[Carollo et al.(2015)]{carollo15} Carollo, D., Beers, T. C., Placco, V. M. et al. 2015, in preparation

\bibitem[Cooper et al.(2010)]{Cooper10} Cooper, A.~P., Cole, S., 
Frenk, C.~S., et al.\ 2010, \mnras, 406, 744 

\bibitem[{{Deason} {et~al.}(2011){Deason}, {Belokurov}, \& {Evans}}]{Deason11b}
{Deason}, A.~J., {Belokurov}, V., \& {Evans}, N.~W. 2011, \mnras, 416, 2903

\bibitem[{{Eggen} {et~al.}(1962){Eggen}, {Lynden-Bell}, \& {andage}}]{Eggen62}
{Eggen}, O.~J., {Lynden-Bell}, D., \& {Sandage}, A.~R. 1962, \apj, 136, 748

\bibitem[Font et al.(2006)]{Font06} Font, A.~S., Johnston, 
K.~V., Bullock, J.~S., \& Robertson, B.~E.\ 2006, \apj, 638, 585 

\bibitem[Jordi(2015)]{jordi15} Jordi, C. 2015, Proceedings of the XI
Scientific Meeting of the Spanish Astronomical Society, A.~J. Cenarro,
F. Figueras, C. Hernandez-Monteagudo, J. Trujillo Bueno, and L.
Valdivielso (eds.), p. 390-401

\bibitem[{{Leaman} {et~al.}(2013){Leaman}, {VandenBerg}, \&
  {Mendel}}]{Leaman13}
{Leaman}, R., {VandenBerg}, D.~A., \& {Mendel}, J.~T. 2013, \mnras, 436, 122

\bibitem[{{Lee} {et~al.}(1994){Demarque} \& {Zinn}}]{Lee1994}
{Lee}, Y.-W., {Demarque}, P., \& {Zinn}, R. 1994, \apj, 423, 248

\bibitem[{{Lee} {et~al.}(2008{\natexlab{a}}){Lee}, {Beers}, {Sivarani}}]{Lee08a}
{Lee}, Y.~S., {Beers}, T.~C., {Sivarani}, T. {\it{et}}~{\it{al}}.
  2008{\natexlab{a}}, \aj, 136, 2022

\bibitem[{{Lee} {et~al.}(2008{\natexlab{b}}){Lee}, {Beers}, {Sivarani}}]{Lee08b}
{Lee}, Y.~S., {Beers}, T.~C., {Sivarani}, T. {\it{et}}~{\it{al}}.
  2008{\natexlab{b}}, \aj, 136, 2050

\bibitem[{{Lee} {et~al.}(2011){Lee}, {Beers}, {Allende Prieto}}]{Lee11}
{Lee}, Y.~S., {Beers}, T.~C., {Allende Prieto}, C. {\it{et}}~{\it{al}}. 2011, \aj, 141, 90

\bibitem[{{McCarthy} {et al.}(2012)}]{mcCarthy2012} 
{McCarthy}, I.G., {Font}, A.S., {Crain}, R.A., {Deason}, A. J., {\it{et}}~{\it{al}}. 2012, \mnras, 420, 2245

\bibitem[{{Paxton} {et~al.}(2011){Paxton}, {Bildsten}, {Dotter}, {Herwig},
  {Lesaffre}, \& {Timmes}}]{paxton:11}
{Paxton}, B., {Bildsten}, L., {Dotter}, A., {Herwig}, F., {Lesaffre}, P., \&
  {Timmes}, F. 2011, \apjs, 192, 3

\bibitem[Perryman et al.(2001)]{perryman01} Perryman, M.~A.~C.,de Boer,
K.~S., Gilmore, G., et al. 2001, \aap, 369, 339

\bibitem[{{Preston} {et~al.}(1991{\natexlab{a}}){Preston}, {Shectman}, \&
  {Beers}}]{Preston91}
{Preston}, G.~W., {Shectman}, S.~A., \& {Beers}, T.~C. 1991{\natexlab{a}},
  \apj, 375, 121

\bibitem[{{Preston} {et~al.}(1991{\natexlab{b}}){Preston}, {Shectman}, \&
  {Beers}}]{Preston91b}
---. 1991{\natexlab{b}}, \apjs, 76, 1001

\bibitem[Reimers(1977)]{reimers:77} Reimers, D. 1977, \aap, 61, 217

\bibitem[{{Santucci} {et~al.}(2015){Santucci}, {Placco}, {Rossi}}]{Santucci15}
{Santucci}, R.~M., {Placco}, V.~M., {Rossi}, S. {\it{et}}~{\it{al}}. 2015, \apj, 801, 116

\bibitem[{{Schlegel} {et~al.}(1998){Schlegel}, {Finkbeiner}, \&
  {Davis}}]{Schlegel98}
{Schlegel}, D.~J., {Finkbeiner}, D.~P., \& {Davis}, M. 1998, \apj, 500, 525

\bibitem[{{Searle} \& {Zinn}(1978)}]{Searle78}
{Searle}, L. \& {Zinn}, R. 1978, \apj, 225, 357

\bibitem[{{Silva Aguirre} {et~al.}(2015){Silva Aguirre}, {Davies}, {Basu}}]{SA15}
{Silva Aguirre}, V., {Davies}, G.~R., {Basu}, S. {\it{et}}~{\it{al}}. 2015, \mnras, 452, 2127

\bibitem[{{Smolinski} {et~al.}(2011){Smolinski}, {Lee}, {Beers}}]{Smolinski11}
{Smolinski}, J.~P., {Lee}, Y.~S., {Beers}, T.~C. {\it{et}}~{\it{al}}. 2011, \aj, 141, 89

\bibitem[{{Soderblom}(2013)}]{Soderblom13}
{Soderblom}, D. 2013, in Asteroseismology of Stellar Populations in the Milky
  Way, 2

\bibitem[Tissera et al.(2012)]{tissera2012} Tissera, P.B., White, S.D.M., 
\& Scannapieco, C. 2012, \mnras, 420, 255

\bibitem[Tissera et al.(2013)]{tissera2013} Tissera, P.B., Scannapieco, C., Beers, T.C., 
\& Carollo, D. 2013, \mnras, 432, 3391

\bibitem[Tumlinson(2010)]{Tumlinson10} Tumlinson, J.\ 2010, \apj, 
708, 1398 

\bibitem[{{VandenBerg} {et~al.}(2012){VandenBerg}, {Bergbusch}, {Dotter}}]{vandenberg:12}
{VandenBerg}, D.~A., {Bergbusch}, P.~A., {Dotter}, A. {\it{et}}~{\it{al}}. 2012, \apj, 755, 15

\bibitem[Vandenberg et al.(2014)]{vandenberg2014} VandenBerg, D.A.,
Bond, H.E., Nelan, E.P., Nissen, P.E., Schaefer, G.H., \& Harmer, D.
2014, \apj, 792, 110

\bibitem[White \& Springel(2000)]{White00} White, S.~D.~M., \& Springel, V.\ 2000, The First Stars, 327 

\bibitem[{{York} {et~al.}(2000){York}, {Adelman}, {Anderson}}]{York00}
{York}, D.~G., {Adelman}, J., {Anderson}, Jr., J.~E. {\it{et}}~{\it{al}}. 2000, \aj, 120, 1579

\bibitem[{{Zhao} \& {Newberg}(2006)}]{Zhao06}
{Zhao}, C. \& {Newberg}, H.~J. 2006, ArXiv:0612034

\bibitem[{{Zinn}(1980)}]{Zinn80}
{Zinn}, R. 1980, \apj, 241, 602

\end{thebibliography}
\end{document}